\documentclass[11pt,onecolumn,a4paper]{article}

\usepackage{amssymb}
\usepackage{graphicx}
\usepackage{url}
\usepackage{float}

\newtheorem{example}{Example}
\newtheorem{theorem}[example]{Theorem}
\newtheorem{corollary}[example]{Corollary}

\newtheorem{lemma}[example]{Lemma}
\newcommand{\qed}{\hfill$\Box$ \vspace{0.5 cm}}
\newenvironment{proof}{{\it Proof: }}{\qed}

\newcommand{\ie}{{\em i.e.}}

\newcommand{\father}{\ensuremath{\mbox{father}}}
\newcommand{\true}{{\em true}}
\newcommand{\false}{{\em false}}
\newcommand{\myindent}{\mbox{\ \ }}

\newcommand{\listing}{\ensuremath{\mbox{\em listing}}}
\newcommand{\pseudolisting}{\ensuremath{\mbox{\em pseudo-listing}}}
\newcommand{\counting}{\ensuremath{\mbox{\em counting}}}
\newcommand{\finding}{\ensuremath{\mbox{\em finding}}}

\newcommand{\algoref}[1]{Algorithm~\ref{algo_#1} ({\em #1})}
\newcommand{\vertexiterator}{{\em vertex-iterator}}
\newcommand{\edgeiterator}{{\em edge-iterator}}

\floatstyle{ruled}
\newfloat{algo}{htb}{algo}
\newcommand{\algorithm}[5]{
\begin{algo}
{\em Input:} #3\\
{\em Output:} #4\\
#5
\caption{{\bf -- {\em #1}.} {\em #2}}
\label{algo_#1}
\end{algo}
}
\floatname{algo}{Algorithm}

\begin{document}

\title{Theory and Practice of Triangle Problems\\
in Very Large (Sparse (Power-Law)) Graphs}
\author{
 Matthieu Latapy\,\footnote{
  LIAFA, CNRS and Universit\'e Paris~7, 2 place
  Jussieu, 75005 Paris, France. latapy@lia\-fa.jus\-sieu.fr} 
 }
\date{}

\maketitle

\begin{abstract}

Finding, counting and/or listing triangles (three vertices
with three edges) in large graphs are natural fundamental problems,
which received recently much attention because of their importance
in complex network analysis. We provide here a detailed state of the
art on these problems, in a unified way. We note that, until now,
authors paid surprisingly little attention to space complexity,
despite its both fundamental and practical interest. We give the
space complexities of known algorithms and discuss their implications.
Then we propose improvements of a known algorithm, as well as a new
algorithm, which are time optimal for triangle
listing and beats previous algorithms concerning
space complexity. They have the additional advantage of performing
better on power-law graphs, which we also study. We finally show with
an experimental study that these two algorithms perform very well in practice,
allowing to handle cases that were previously out of reach.

\end{abstract}

\section{Introduction.}
\label{sec-intro}

A {\em triangle} in an undirected graph is a set of three vertices
such that each possible edge between them is present in the graph.
Following classical conventions, we call \finding, \counting\ 
and \listing\ the problems of deciding if a given graph contains any
triangle, counting the number of triangles in the graph,
and listing all of them, respectively. We moreover
call \pseudolisting\ the problem of counting {\em for each vertex} the
number of triangles to which it belongs. We refer to all these problems
as a whole by {\em triangle problems}.

Triangle problems may be considered as classical,
natural and fundamental algorithmic questions, and have been studied as such
\cite{itai78finding,chiba85arboricity,alon94finding,alon97finding,schank05finding,schank05findingWEA}.

Moreover, they gained recently much practical importance since they are central
in so-called {\em complex network analysis}, see for instance~\cite{watts98collective,brandes05lncs,albert02statistical,eubank04structural}.
First, they are involved in the
computation of one of the main statistical property used to describe large
graphs met in practice, namely the clustering coefficient \cite{watts98collective}.
The clustering coefficient of a vertex $v$ (of degree at least $2$) is the probability
that any two randomly chosen neighbors of $v$ are linked together. It is computed by
dividing the number of triangles containing $v$ by the number of possible
edges between its neighbors, \ie\ $d(v) \choose 2$ if $d(v)$ denotes the number of
neighbors of $v$.
One may then define the clustering coefficient of the whole graph as the average of
this value for all the vertices (of degree at least $2$). Likewise, the {\em transitivity
ratio}\,\footnote{
Even though some authors make no distinction between the two notions, they are different,
see for instance \cite{bollobas02mathematical,schank04approximating}. Both have their
own advantages and drawbacks,
but discussing this is out of the scope of this contribution.}
\cite{harary57toward,harary79matrix} is defined as
$\frac{3\cdot N_{\Delta}}{N_{\vee}}$
where $N_{\Delta}$ denotes the number of triangles in the graph and
$N_{\vee}$ denotes the number of
connected triples, \ie\ sets of three vertices with at least two edges,
in the graph.

In the context of complex network analysis, triangles also play a key
role in the study of motif occurrences, \ie\ the presence of special
(small) subgraphs in given (large) graphs. This has been studied in particular
in protein interaction networks, where some motifs may correspond to biological
functions, see for instance \cite{milo02network,yeger04network}. Triangles often
are building blocks of these motifs.

\medskip

Finally, triangle finding, counting, pseudo-listing and/or listing appear
as key issues both from a fundamental point of view and for
practical purpose. The aim of this contribution is to review the algorithms
proposed until now for solving these problems with both a fundamental
perspective (we discuss asymptotic complexities and give detailed proofs)
and a practical one (we discuss space requirements and graph encoding,
and we evaluate algorithms with some experiments).

We note that, until now, authors paid surprisingly little attention to space
requirements of their algorithms for triangle problems; this however is an
important limitation in practice, and this also induces interesting theoretical
questions. We will therefore discuss this (all space complexity results stated
in this paper are new, though very simple in most cases), and we will propose
space-efficient algorithms.

The paper is organised as follows. After a few preliminaries
(Section~\ref{sec-prelim}), we begin with results on finding, counting and pseudo-listing
problems, between which basically no difference in complexity is known (Section~\ref{sec-first}).
Then we turn to the harder problem of triangle listing, in Section~\ref{sec-listing}.
In these parts of the paper, we deal with both the general case (no assumption is
made on the graph) and on the important case where the graph is sparse.
Many very large graphs met in practice also have
heterogeneous degrees; we focus on this case in Section~\ref{sec-pl}. Finally,
we present experimental evaluations in Section~\ref{sec-exp}. We summarise the
current state of the art and we point out the main perspectives in
Section~\ref{sec-conclu}.

\section{Preliminaries.}
\label{sec-prelim}

Throughout the paper, we consider an undirected\,\footnote{\ie\ we make no
difference between $(u,v)$ and $(v,u)$ in $V \times V$.} graph $G=(V,E)$
with $n = |V|$ vertices and $m = |E|$ edges. We suppose that
$G$ is simple ($(v,v)\not\in E$ for all $v$, and there is no multiple
edge). We also assume that $m \in \Omega(n)$; this is a classical convention which
plays no role in our algorithms but makes complexity formulae simpler.
We denote by $N(v)\ =\ \{u\in V,\ (v,u)\in E\}$ the neighborhood of $v \in V$ and
by $d(v) = |N(v)|$ its degree. We also denote by 
$d_{\max}$ the maximal degree in $G$: $d_{\max} = \max_v\{d(v)\}$.

Before entering in the core of this paper, we need to discuss a few issues
that will play an important role in the following. They are necessary to
make the discussion all along the paper precise and rigorous.

\subsubsection*{Graph encodings.}

First note that we will always suppose that the graph is stored
in central memory\,\footnote{Approaches not requiring this, based on streaming
algorithms for instance \cite{henzinger98computing,bar02reduction,jowhari05new},
or various methods to compress the graph \cite{boldi04www,boldi04dcc}, also exist.
This is however out of the scope of this paper.}.
There are basically two ways to do this:
\begin{itemize}
\item
$G$ may be encoded by its adjacency matrix $A$ defined by
$A_{ij}=1$ if $(i,j) \in E$, $A_{ij}=0$ else. This has a $\Theta(n^2)$
space cost. Since $m$ may be up to $\Theta(n^2)$,
this representation is space optimal in this case (but it is not as soon
as the graph is sparse, \ie\ $m\in o(n^2)$), and makes it possible to
test the presence of any edge in $\Theta(1)$. Note however that
one cannot run through $N(v)$
in $O(d(v))$ time with such a representation: one needs $\Theta(n)$ time.
Since $d(v)$ may be up to $\Theta(n)$, this is not a problem in the general case.
\item
$G$ may be encoded by a {\em simple compact representation}: for each
vertex $v$ we can access the set of its neighbors $N(v)$ and
its degree $d(v)$ in $\Theta(1)$
time and space cost. The set $N(v)$ usually is encoded using a linked list
or an array, in order to be
able to run through it in $\Theta(d(v))$ time and $\Theta(1)$ space.
It may moreover be sorted (an order on the vertices is supposed to be
given). This representation has the advantage of being space efficient:
it needs only $\Theta(m)$ space. However, testing the presence
of the edge $(u,v)$ is in $\Theta(d(v))$ time ($O(\log(d(v)))$ if
$N(v)$ is a sorted array). We call any representation having these
properties a {\em simple compact representation} of $G$.
\end{itemize}

Since the basic operations of such representations do not have the
same complexity, they may play a key role in algorithms using them.
We will see that this is indeed the case in our context. We note
moreover that, in the context of large graph manipulation, the
adjacency matrix often is untractable because of its space requirements.
This is why one generally uses (sorted) simple compact representations
in practice.

One may easily convert any simple compact
representation of $G$ into its adjacency array representation,
in time $\Theta(m)$ using $\Theta(n)$ additional space (it suffices
to transform iteratively each set $N(v)$ and to free the memory
used by the previous representation at each step).
Moreover, once the adjacency array representation of $G$ is available,
one may compute its sorted version in 
$\Theta\left(\sum_v d(v)\cdot\log(d(v))\right)
\ \subseteq\ O\left(\sum_v d(v)\cdot\log(n)\right)
\ =\ O(m\cdot\log(n))$
time and $\Theta(1)$ additional space.
One may therefore intuitively make no difference between any simple
compact representation of $G$ and its sorted adjacency array
representation, as long as the overall algorithm complexity is in
$\Omega(m\cdot\log(n))$ time and $\Omega(n)$ space.

One may also obtain a simple compact representation of $G$ from
its adjacency matrix in time $\Theta(n^2)$ and additional space
$\Theta(n)$ (provided that one does not need the matrix anymore,
else it costs $\Theta(m)$). This cost is not neglectible in most
cases, and thus we will suppose that algorithms that need the two
representations receive them both as inputs.

Finally, note that one may use more subtle structures to encode
the sets $N(v)$ for all $v$. Balanced trees and hashtables are the
most classical ones. Since we focus on worst case analysis (see below),
such encodings have no impact on our results, and so we make
no difference between them and any other simple compact representation.

\subsubsection*{(Additional) Space complexity.}

As explained above, storing the graph itself generally is
in $\Theta(n^2)$ or $\Theta(m)$ space complexity.
Moreover, the space requirements of the algorithms we will study
are, in most cases, lower than the space requirements of the graph
storing. Therefore, their space complexity is the one of the chosen
graph representation, which makes little sense.

However, limiting the space needed by the algorithm {\em
in addition} to the one needed to store the graph often is a key issue in
practice: current main limitation in triangle problems on real-world complex
networks is space requirements. We illustrate this in Section~\ref{sec-exp}.

For these reasons, the space complexities we discuss concern the
{\em additional} space needed by the algorithm, \ie\ not including the
graph storage. As we will see, this notion makes a significant difference
between various algorithms, and therefore also has a fundamental interest.

Likewise, and following classical conventions, we do not include the size
of the output in our space complexities. Otherwise, triangle listing would
need $\Omega(n^3)$ space in the worst case, and pseudo-listing would need
$\Omega(n)$ space, which brings little information, if any.

\subsubsection*{Worst case complexity, and graph families.}

All the complexities we discuss in this paper are {\em worst case}
complexities, in the sense that they are bounds for the time and space
needs of the algorithms, on any input. In
most cases, these bounds are tight (leading to the use of the $\Theta()$
notation, see for instance \cite{cormen01book} for definitions).
In other words, we say that an algorithm is in $\Theta(f(n))$ if
there exists
an instance of the input such that the algorithm runs with this
complexity (even if some instances induce lower complexity).
In several case, however, the worst case complexity actually is
the complexity for any input (in the case of Theorem~\ref{th-direct},
for instance, and for most space complexities).

It would also be of high
interest to study the expected behavior of triangle algorithms, in addition to the
worst case one. This has
been done in some cases; for instance, it is proved in \cite{itai78finding}
that \vertexiterator\ (see Section~\ref{sec-listing-basic}) has expected time complexity in
$O(n^{\frac{5}{3}})$. Obtaining such results however often is very difficult, and their
relevance for practical purpose is not always clear: the choice of a
model for the average input is a difficult task (in our context, random
graphs would be an unsatisfactory choice \cite{brandes05lncs,albert02statistical,watts98collective}).
We therefore
focus on worst case analysis, which has the advantage of giving guarantees
on the behaviors of algorithms, on any input.

Another interesting approach is to study (worst case) complexities on
given graph families. This has already been done on various cases, the
most important ones probably being the sparse graphs,
\ie\ graphs in which $m$ is in $o(n^2)$.
This is motivated by the fact that most real-world complex networks lead to
such graphs, see for instance \cite{brandes05lncs,albert02statistical,watts98collective}.
In general, it is even assumed that $m$ is in $O(n)$. Recent studies however
show that, despite the fact that $m$ is small compared to $n^2$, it may be in $\omega(n)$
\cite{leskovec05densification,holme2003dating,latapy06measuring}.
Other classes of graphs have been considered, like for instance planar
graphs: it is shown in \cite{itai78finding} that
one may decide if any planar graph contains a triangle in $O(n)$ time.

We do not detail all these results here. Since we are particularily
interested in real-world complex networks, we present in detail
the results concerning sparse graphs all along the paper. We also
introduce new results on power-law graphs (Section~\ref{sec-pl}), which
capture an important property met in practice. A survey on available
results on specific classes of graphs remains to be done, and is out of
the scope of this paper.

\section{The fastest algorithms for finding, counting, and pseudo-listing.}
\label{sec-first}

The fastest algorithm known for pseudo-listing relies on fast
matrix product \cite{itai78finding,alon94finding,alon97finding,coppersmith90matrix}.
Indeed, if one considers the adjacency matrix $A$ of $G$ then the value $A^3_{vv}$
on the diagonal of $A^3$ is nothing but twice the number of triangles to which $v$
belongs, for any $v$. Finding, counting and pseudo-listing triangle problems can
therefore be solved in $O(n^\omega)$ time, where $\omega<2.376$ is the
fast matrix product exponent \cite{coppersmith90matrix}. This was first noticed
in 1978 \cite{itai78finding}, and currently no faster algorithm is known for any
of these problems in the general case, even for triangle finding (but this is no longer
true when the graph is sparse, see Theorem~\ref{th-ayz-pseudo-listing} below).

This approach naturally needs the graph to be given by its adjacency matrix
representation. Moreover, it makes it necessary to compute and store the matrix
$A^2$, leading to a $\Theta(n^2)$ space complexity in addition to the adjacency
matrix storage.

\begin{theorem}[\cite{itai78finding,coppersmith90matrix}]
Given the adjacency matrix representation of $G$, it is possible to solve
triangle finding, counting and pseudo-listing in
$O(n^{\omega}) \subset O(n^{2.376})$ time and $\Theta(n^2)$ space on
$G$ using fast matrix product.
\label{th-matrix}
\end{theorem}

This time complexity is the current state of our knowledge, as long as one makes
no assumption on $G$. Note that no lower bound is known for this complexity; therefore
faster algorithms may be designed.

As we will see, there exists (slower) algorithms with lower space complexity
for these problems. Some of these algorithms only need a simple compact representation
of $G$. They are derived from listing algorithms, which we
present in Section~\ref{sec-listing}.

\medskip

One can design faster algorithms if $G$ is sparse.
In \cite{itai78finding}, it was first proved that triangle finding, counting, pseudo-listing and listing\,\footnote{
The original results actually concern triangle {\em finding}
but they can easily be extended to counting, pseudo-listing and listing at no cost;
we present such an extension in Section~\ref{sec-listing},
\algoref{tree-listing}.}
can be solved in $\Theta(m^{\frac{3}{2}})$ time and $\Theta(m)$ space.
This result has been improved in \cite{chiba85arboricity} using a property of the graph (namely
arboricity) but the worst case complexites were unchanged. No better
result was known until 1995 \cite{alon97finding,alon94finding}, where the authors prove
Theorem~\ref{th-ayz-pseudo-listing} below\,\footnote{
Again, the original results concerned triangle {\em finding}, but may easily be
extended to pseudo-listing, see \algoref{ayz-pseudo-listing}, and listing,
see \algoref{ayz-listing}. This was first proposed in \cite{schank05finding,schank05findingWEA}.
These algorithms have also been generalized to longer cycles in
\cite{yuster04detecting} but this is out of the scope of this paper.},
which constitutes a significant improvement although it relies
on very simple ideas. We detail the proof and give a slightly different version,
which will be useful in the following (similar ideas are used in Section~\ref{sec-listing-compact},
and this proof permits a straightforward extension of this theorem in
Section~\ref{sec-pl}).

\algorithm{ayz-pseudo-listing}{Counts for all $v$ the triangles in $G$ containing $v$ \cite{alon97finding,alon94finding}.}{
any simple compact representation of $G$, its adjacency matrix $A$, and an integer $K$}{
$T$ such that $T[v]$ is the number of triangles in $G$ containing $v$}{
1. initialise $T[v]$ to $0$ for all $v$\\
2. for each vertex $v$ with $d(v)\le K$:\\
\myindent 2a. for each pair $\{u,w\}$ of neighbors of $v$:\\
\myindent\myindent 2aa. if $A[u,w]$ then:\\
\myindent\myindent\myindent 2aaa. increment $T[v]$\\
\myindent\myindent\myindent 2aab. if $d(u)>K$ and $d(w)>K$ then increment $T[u]$ and $T[w]$\\
\myindent\myindent\myindent 2aac. else if $d(u)>K$ and $u>v$ then increment $T[u]$\\
\myindent\myindent\myindent 2aad. else if $d(w)>K$ and $w>v$ then increment $T[w]$\\
3. let $G'$ be the subgraph of $G$ induced by $\{v,\ d(v)>K\}$\\
4. construct the adjacency matrix $A'$ of $G'$\\
5. compute $A'^3$ using fast matrix product\\
6. for each vertex $v$ with $d(v) > K$:\\
\myindent 6a. add to $T[v]$ half the value in $A'^3_{vv}$
}

\begin{theorem}[\cite{alon97finding,alon94finding}]
Given any simple compact representation of $G$ and its adjacency matrix,
it is possible to solve triangle finding, counting and pseudo-listing
on $G$ in
$O(m^{\frac{2\cdot\omega}{\omega+1}}) \subset O(m^{1.41})$ time and
$\Theta\left(m^{\frac{4}{\omega+1}}\right) \subset O(m^{1.185})$ space;
\algoref{ayz-pseudo-listing}\ achieves this if one takes $K \in \Theta(m^{\frac{\omega-1}{\omega+1}})$.
\label{th-ayz-pseudo-listing}
\end{theorem}
\begin{proof}
Let us first show that \algoref{ayz-pseudo-listing}\ solves pseudo-listing
(and thus counting and finding). Consider a triangle in $G$ that contains a
vertex with degree at most $K$; then it is discovered in lines~2a
and~2aa. Lines~2aaa to~2aad ensure that it is counted exactly once for each
vertex it countains. Consider now the triangles in which all the vertices have
degree larger than $K$. Each of them induces a triangle in $G'$, and $G'$ contains
no other triangle. These triangles are counted
using the matrix product approach (lines~5,~6 and~6a), and finally all the triangles in $G$ are
counted for each vertex.

Let us now study the time complexity of \algoref{ayz-pseudo-listing}\ in function
of $K$. For each vertex $v$
with $d(v)\le K$, one counts the number of triangles containing $v$ in
$\Theta(d(v)^2) \subseteq O(d(v)\cdot K)$ thanks to the simple compact representation
of $G$.
If we sum over all the vertices in the graph this leads to
a time complexity in $O(m.K)$ for lines~2 to~2aad.
Now notice that there cannot be more than $\frac{2\cdot m}{K}$ vertices $v$ with $d(v)>K$.
Line~4
constructs (in $O\left(m+(\frac{m}{K})^2\right)$ time, which plays no role in the
global complexity) the adjacency matrix of the subgraph $G'$ of $G$ induced by these
vertices. Using fast matrix
product, line~5 computes the number of triangles for each vertex in $G'$ in
time $O\left(\left(\frac{m}{K}\right)^{\omega}\right)$. Finally, we obtain the
overall time complexity of the algorithm:
$O\left(m.K + \left(\frac{m}{K}\right)^{\omega}\right)$.

In order to minimize this, one has to search for a value of $K$ such that
$m\cdot K \in \Theta((\frac{m}{K})^\omega)$. This leads to
$K\in \Theta(m^{\frac{\omega-1}{\omega+1}})$, which gives the announced time
complexity.

Concerning space complexity, the key point is that one has to construct $A'$, $A'^2$
and $A'^3$. The matrix $A'$ may contain
$\frac{2\cdot m}{K}$ vertices, leading to a
$\Theta\left(\left(\frac{m}{K}\right)^2\right) =
\Theta\left(m^{2\cdot\left(1-\frac{\omega-1}{\omega+1}\right)}\right) =
\Theta\left(m^{\frac{4}{\omega+1}}\right)$ space complexity.
\end{proof}

Note that one may also use {\em sparse} matrix product algorithms, see for
instance \cite{yuster04fast}. However, the
matrix $A^2$ may not be sparse (in particular if there are vertices with
large degrees, which is often the case in practice as discussed in
Section~\ref{sec-pl}). But algorithms may take benefit from the fact
that {\em one} of the two matrices involved in a product is sparse, and
there also exists algorithms for products of more than two sparse
matrices. These approaches lead to algorithms
whose efficiency depends on the exact relation between $m$ and $n$: it
depends on the relation between $n$ and $m$ which algorithm is the fastest.
Discussing this further therefore is quite complex, and it is out of
the scope of this paper.

\medskip

In conclusion, despite the fact that the algorithms presented in this section are
asymptotically very fast, they have two important limitations. First, they have
a prohibitive space cost, since the matrices involved in the computation
(in addition to the adjacency matrix, but it is considered as the
encoding of $G$ itself) may need $\Theta(n^2)$ space. Moreover, the
fast matrix product
algorithms are quite intricate, which leads to difficult implementations
with high risks of errors. This also leads to large constant factors in
the complexities, which have no importance at the asymptotic limit
but may play a significant role in practice.

For these reasons, and despite the fact that they clearly are of prime theoretical
importance, these algorithms have limited practical impact. Instead, one
generally uses one of the {\em listing} algorithms (adapted accordingly)
that we detail now.

\section{Time-optimal listing algorithms.}
\label{sec-listing}

First notice that there may be ${n \choose 3} \in \Theta(n^3)$ triangles
in $G$. Likewise, there may be $\Theta(m^{\frac{3}{2}})$ triangles, since $G$ may
be a clique of $\sqrt{m}$ vertices (thus containing
${\sqrt{m} \choose 3} \in \Theta(m^{\frac{3}{2}})$ triangles). This gives the
following lower bounds for the time complexity of any triangle listing
algorithm.

\begin{lemma}[\cite{itai78finding,schank05finding,schank05findingWEA}]
Listing all triangles in $G$ is in $\Omega(n^3)$ and $\Omega(m^{\frac{3}{2}})$ time.
\label{lem-bound}
\end{lemma}

In this section, we first observe that the time complexity $\Theta(n^3)$
can easily be reached (Section~\ref{sec-listing-basic}). However, $\Theta(m^{\frac{3}{2}})$ is much better in the
case of sparse graphs. We present more subtle algorithms that reach
this bound (Section~\ref{sec-listing-sparse}). Again, space complexity is a key issue, and we discuss
this for each algorithm. We will see that algorithms proposed until
now either rely on the use of adjacency matrices and/or have a
$\Omega(m)$ space complexity. We improve this by
proposing algorithms that reach a $\Theta(n)$ space complexity, while
needing only a simple compact representation of $G$, and still in
$\Theta(m^{\frac{3}{2}})$ time (Section~\ref{sec-listing-compact}).

\subsection{Basic algorithms.}
\label{sec-listing-basic}

One may trivially obtain a listing
algorithm in $\Theta(n^3)$ (optimal) time with the matrix representation of $G$ by
testing in $\Theta(1)$ time any possible triple of vertices. Moreover, this
algorithm has the optimal space complexity $\Theta(1)$.

\begin{theorem}[\cite{schank05finding,schank05findingWEA} and folklore]
Given the adjacency matrix representation of $G$, it is possible to solve
triangle listing in $\Theta(n^3)$ time and $\Theta(1)$ space
using the direct testing of every triple of vertices.
\label{th-direct}
\end{theorem}

This approach however has severe drawbacks. First, it needs the adjacency
matrix of $G$. More importantly, its complexity does not depend on the
actual properties of $G$; it always needs $\Theta(n^3)$ computation
steps even if the graph contains very few edges. It must however be clear
that, if almost all triples of vertices form a triangle, no better asymptotic
bound can be attained, and the simplicity of this algorithm makes it very
efficient in these cases.

\medskip

In order to obtain faster algorithms on sparse graphs, while keeping the
implementation very simple, one often uses the following algorithms. The
first one, introduced in \cite{itai78finding} and called
\vertexiterator\ in \cite{schank05finding,schank05findingWEA}, consists in
iterating \algoref{vertex-listing}\ on each vertex of $G$.
The second one, which seems to be the most widely used
algorithm\,\footnote{It is for instance implemented in the widely used
complex network analysis software {\em Pajek} \cite{batagelj01subquadratic,batagelj98pajek,vlado06perso}.},
consists in iterating \algoref{edge-listing}\ over each edge in $G$.
It was also first introduced 
in \cite{itai78finding}, and discussed in \cite{schank05finding,schank05findingWEA}
where the authors call it \edgeiterator.

\algorithm{vertex-listing}{Lists all the triangles containing a given vertex \cite{itai78finding}.}{
any simple compact representation of $G$, its adjacency matrix $A$, and a vertex $v$}{
all the triangles to which $v$ belongs}{
1. for each pair $\{u,w\}$ of neighbors of $v$:\\
\myindent 1a. if $A_{uw} = 1$ then output triangle $\{u,v,w\}$
}

\algorithm{edge-listing}{Lists all the triangles containing a given edge \cite{itai78finding}.}{
any sorted simple compact representation of $G$, and an edge $(u,v)$ of $G$}{
all the triangles in $G$ containing $(u,v)$}{
1. for each $w$ in $N(u)\cap N(v)$:\\
\myindent 1a. output triangle $\{u,v,w\}$
}

\begin{theorem}[\cite{itai78finding,schank05finding,schank05findingWEA}]
Given any simple compact representation of $G$ and its adjacency matrix,
it is possible to list all its triangles in
$\Theta\left(\sum_v d(v)^2\right)$, $\Theta(m\cdot d_{\max})$, $\Theta(m\cdot n)$, and $\Theta(n^3)$
time and $\Theta(1)$ space;
\vertexiterator\ achieves this.
\label{th-vertex-listing}
\end{theorem}
\begin{proof}
The fact that \algoref{vertex-listing}\ list all the triangles to which a vertex $v$ belongs
is straightforward. Then, iterating over all vertices gives three times each triangle; if one
wants each triangle only once it is sufficient to restrict the output of triangles to
the ones for which $\eta(w)>\eta(v)>\eta(u)$, for any injective numbering $\eta()$ of the vertices.

Thanks to the simple compact representation of $G$, the pairs of
neighbors of $v$ may be computed in $\Theta(d(v)^2)$ time and $\Theta(1)$ space (this
would be impossible with the adjacency matrix only). Thanks to the adjacency matrix,
the test in line~1a may be processed in $\Theta(1)$ time and space (this
would be impossible with the simple compact representaton only).
The time complexity of \algoref{vertex-listing}\ therefore is in $\Theta(d(v)^2)$ time
and $\Theta(1)$ space. The $\Theta(\sum_v d(v)^2)$ time and $\Theta(1)$ space
complexity of the overall algorithm follows.
Moreover, we have $\Theta(\sum_v d(v)^2) \subseteq O(\sum_v d(v)\cdot d_{\max}) = O(m\cdot d_{\max}) \subseteq O(m\cdot n) \subseteq O(n^3)$,
and all these complexity may be attained in the worst case (clique of $n$ vertices), hence the results.
\end{proof}

\begin{theorem}[\cite{itai78finding,schank05finding,schank05findingWEA} and folklore]
Given any sorted simple compact representation of $G$, it is possible to
list all its triangles in
$\Theta(m\cdot d_{\max})$, $\Theta(m\cdot n)$ and $\Theta(n^3)$ time and $\Theta(1)$ space;
The \edgeiterator\ algorithm achieves this.
\label{th-edge-listing}
\end{theorem}
\begin{proof}
The correctness of the algorithm is immediate. One may proceed like in the proof
of Theorem~\ref{th-vertex-listing} to obtain each triangle only once.

Each edge $(u,v)$ is treated in time $\Theta(d(u)+d(v))$ (because $N(u)$ and $N(v)$
are sorted) and $\Theta(1)$ space.
We have $d(u)+d(v) \in \Theta(d_{\max})$, therefore the overall complexity
is in $O(m\cdot d_{\max}) \subseteq O(m\cdot n) \subseteq O(n^3)$. In the worst
case (clique of $n$ vertices) all these complexity are tight.
\end{proof}

First note\,\footnote{We also note that another $O(m\cdot n)$ time algorithm
was proposed in \cite{monien85how} for a more general problem. In the case
of triangles, it does not improve \vertexiterator\ and \edgeiterator, which
are much simpler,
therefore we do not detail it here.} that these algorithms are optimal in the worst case, just like
the direct method (Lemma~\ref{lem-bound} and Theorem~\ref{th-direct}). However, there are much more efficient on sparse graphs,
in particular if the maximal degree is low \cite{batagelj01subquadratic}, since
they both are in $\Theta(m\cdot d_{\max})$ time. If the maximal degree is a
constant, \vertexiterator\ even is in $\Theta(n)$ time.
Moreover, both algorithms only need $\Theta(1)$ space, which makes them very
interesting from this perspective (we will see that there is no known faster
algorithm with this space requirement).

However, \vertexiterator\ has a severe drawback: it needs the adjacency matrix
of $G$ {\em and} a simple compact representation. Instead, \edgeiterator\ only
needs a sorted simple compact representation, which is often available in
practice\,\footnote{Recall that one may sort the
simple compact representation of $G$ in $\Theta(m\log(n))$ time and $\Theta(n)$ space,
if needed.}. Moreover, \edgeiterator\ runs in $\Theta(1)$
space, which makes it very compact. Because of these two reasons, and because
of its simplicity, it is widely used in practice.

The performance of these algorithms however are quite poor when the
maximal degree is unbounded, and in particular if it grows like $n$.
They may even be asymptotically sub-optimal on sparse graphs
and/or on graphs with
some vertices of high degree, which often appear in practice (we 
discuss this further in Section~\ref{sec-pl}). It is however possible
to design time-optimal listing algorithms for sparse graphs, which
we detail now.

\subsection{Time-optimal listing algorithms for sparse graphs.}
\label{sec-listing-sparse}

Several algorithms have been proposed that reach the $\Theta(m^{\frac{3}{2}})$ bound
of Lemma~\ref{lem-bound}, and thus are
time optimal on sparse graphs (note that this is also optimal for dense
graphs, but we have seen in Section~\ref{sec-listing-basic} much simpler algorithms for
these cases). Back in 1978, an algorithm was proposed to
{\em find} a triangle in $\Theta(m^{\frac{3}{2}})$ time and $\Theta(m)$ space \cite{itai78finding}.
Therefore it is slower than the ones discussed in Section~\ref{sec-first} for finding, but
it may be extended to obtain a {\em listing} algorithm with the same complexity.
We first present this below. Then,
we detail two simpler solutions with this complexity, proposed
recently in \cite{schank05finding,schank05findingWEA}.
The first
one consists in a simple extension of \algoref{ayz-pseudo-listing}; the other
one, named {\em forward}, has the advantage of being very efficient in
practice \cite{schank05finding,schank05findingWEA}. Moreoever, we show
in Section~\ref{sec-listing-compact} that it may be slightly modified to
reach a $\Theta(n)$ space cost.

\subsubsection*{An approach based on covering trees \cite{itai78finding}.}

We use here the classical notions of covering trees and connected components,
as defined for instance in \cite{cormen01book}. Since they are very classical, we do not
recall them. We just note that a covering tree of each connected component of
any graph may be computed in time linear in the number of edges of this graph,
and space linear in its number of vertices (typically using a breadth-first
search). One then has access to the father of any vertex in $\Theta(1)$
time and space.

In \cite{itai78finding}, the authors propose a triangle {\em finding} algorithm
in $\Theta(m^{\frac{3}{2}})$ time and $\Theta(m)$ space. We present here a simple
extension of this algorithm to solve triangle {\em listing} with the same
complexity. To achieve this, we need the following lemma, which is a simple
extension of Lemma~4 in~\cite{itai78finding}.

\begin{lemma}[\cite{itai78finding}]
Let us consider a covering tree for each connected component of $G$, and a
triangle $t$ in $G$ having an edge in one of these trees.
Then there exists an edge $(u,v)$ in $E$ but in none of these trees, such that
$t=\{u,v,\father(v)\}$.
\label{lem-tree}
\end{lemma}
\begin{proof}
Let $t = \{x,y,z\}$ be a triangle in $G$, and let $T$ be the tree that contains
an edge of $t$. We can suppose without loss of generality that this edge is
$(x,y=\father(x))$.
Two cases have to be considered. First, if $(x,z) \not\in T$ then
it is in none of the trees, and taking $v=x$ and $u=z$ satisfies the claim. Second, if
$(x,z) \in T$ then we have $\father(z)=x$ (because $\father(x) = y \not= z$).
Moreover, $(y,z)\not\in T$ (else $T$ would contain a cycle, namely $t$).
Therefore taking $v=z$ and $u=y$ satisfies the claim.
\end{proof}

\algorithm{tree-listing}{Lists all the triangles in a graph \cite{itai78finding}.}{
any simple compact representation of $G$, and its adjacency matrix $A$}{
all the triangles in $G$}{
1. while there remains an edge in $E$:\\
\myindent 1a. compute a covering tree for each connected component of $G$\\
\myindent 1b. for each edge $(u,v)$ in none of these trees:\\
\myindent\myindent 1ba. if $(\father(u),v) \in E$ then output triangle $\{u,v,\father(u)\}$\\
\myindent\myindent 1bb. else if $(\father(v),u) \in E$ then output triangle $\{u,v,\father(v)\}$\\
\myindent 1c. remove from $E$ all the edges in these trees
}

This lemma shows that, given a covering tree of each connected component of
$G$, one may find triangles by checking for each edge $(u,v)$ that belongs
to none of these trees if $\{u,v,\father(v)\}$ is a triangle. Then, all the
triangles containing $(v,\father(v))$ are discovered. This leads to
\algoref{tree-listing}, and to
the following result (which is a direct extension of the one concerning
triangle {\em finding} described in \cite{itai78finding}).
\begin{theorem}[\cite{itai78finding}]
Given any simple compact representation of $G$ and its adjacency matrix,
it is possible to list all its triangles in $\Theta(m^{\frac{3}{2}})$ time
and $\Theta(n)$ space; \algoref{tree-listing}\ achieves this.
\end{theorem}
\begin{proof}
Let us first prove that the algorithm is correct. It is clear that the
algorithm may only output triangles. Suppose that one is missing. But
all its edges have been removed when the computation stops, and so
(at least) one
of its edges was in a tree at some step. Let us consider the first
such step (therefore the three edges of the triangle are present).
Lemma~\ref{lem-tree}
says that there exists an edge satisfying the condition tested in lines~1b
and~1ba, and thus the triangle was discovered at this step. Finally, we
reach a contradiction, and thus all
triangles have been discovered.

Now let us focus on the time complexity. Following \cite{itai78finding},
let $c$ denote the number of connected components at the current step of
the algorithm. The value of $c$ increases during the computation, until
it reaches $c=n$. Two cases have to be considered. First suppose that
$c \le n-\sqrt{m}$. During this step of the algorithm, $n-c \ge n-(n-\sqrt{m}) = \sqrt{m}$
edges are removed. And thus there can be no more than $\frac{m}{\sqrt{m}} = \sqrt{m}$
such steps. Consider now the other case, $c > n-\sqrt{m}$. The maximal
degree then is at most $n - c < n -(n-\sqrt{m}) = \sqrt{m}$, and,
since the degree of each vertex (of non-null degree) decreases at each
step, there can be no more than $\sqrt{m}$ such steps. Finally, the total
number of steps is bounded by $2\cdot \sqrt{m}$. Moreover, each step costs
$O(m)$ time: the test in line~1ba is in $\Theta(1)$ time thanks to the 
adjacency matrix, and line~1b finds the $O(m)$ edges on which it is
ran in $O(m)$ time thanks to the $\father()$ relation which is in
$\Theta(1)$ time. This leads to the $O(m^{\frac{3}{2}})$ time complexity,
and, from Lemma~\ref{lem-bound}, this bound is tight.

Finally, let us focus on the space complexity. Suppose that removing an
edge $(u,v)$ is done by setting $A_{uv}$ and $A_{vu}$ to $0$, but without changing
the simple compact representation. Then, the actual presence of an edge
in the simple compact representation can be tested with only a constant
additional cost by checking that the corresponding entry in the matrix
is equal to $1$. Therefore,
this way of removing edges induces no significant additional time cost,
while allowing a computation in $\Theta(n)$ space (needed for the trees).
\end{proof}

The space complexity obtained here is very good (and we will see that
we are unable to obtain better ones), but it relies on the fact that
the graph is given both in its adjacency matrix representation {\em and}
a simple compact one. This reduces significantly the practical relevance
of this approach concerning reduced space complexity. We will
see in the next section algorithms that have the same time and space
complexities but needing only a simple compact representation of $G$.

\subsubsection*{An extension of \algoref{ayz-pseudo-listing} \cite{alon97finding,alon94finding,schank05finding,schank05findingWEA}.}

The fastest known algorithm for finding, counting, and pseudo-listing
triangles, namely \algoref{ayz-pseudo-listing}, was proposed in \cite{alon97finding,alon94finding} and we
detailed it in Section~\ref{sec-first}. As proposed first in
\cite{schank05finding,schank05findingWEA}, it is easy to modify it
to obtain a {\em listing} algorithm, namely \algoref{ayz-listing}.

\algorithm{ayz-listing}{Lists all the triangles in a graph \cite{alon97finding,alon94finding,schank05finding,schank05findingWEA}.}{
any simple compact representation of $G$, its adjacency matrix $A$, and an integer $K$}{
all the triangles in $G$}{
1. for each vertex $v$ with $d(v)\le K$:\\
\myindent 1a. output all triangles containing $v$ with \algoref{vertex-listing}, without duplicates\\
2. let $G'$ be the subgraph of $G$ induced by $\{v,\ d(v)>K\}$\\
3. compute a sorted simple compact representation of $G'$\\
4. list all triangles in $G'$ using \algoref{edge-listing}
}

\begin{theorem} \cite{schank05finding,schank05findingWEA,alon97finding,alon94finding}
Given any simple compact representation of $G$ and its adjacency matrix, it is possible to list
all its triangles in $\Theta(m^{\frac{3}{2}})$ time and
$\Theta(m)$ space; \algoref{ayz-listing}\ achieves this if one takes $K \in \Theta(\sqrt{m})$.
\label{th-ayz-listing}
\end{theorem}
\begin{proof}
First recall that one may sort the simple compact representation
of $G$ in $O(m\cdot\log(n))$ time and $\Theta(1)$ space. This has
no impact on the overall complexity of \algoref{ayz-listing}, thus
we suppose in this proof that the representation is sorted.

In a way similar to the proof of Theorem~\ref{th-ayz-pseudo-listing} let us first
express the complexity of \algoref{ayz-listing}\ in terms of $K$.
Using the $\Theta(d(v)^2)$ complexity of \algoref{vertex-listing}\ we obtain
that lines~1 and~1a have a cost in $O(\sum_{v, d(v)\le K} d(v)^2) \subseteq
O(\sum_{v, d(v)\le K} d(v)\cdot K \subseteq O(m\cdot K)$ time. Moreover, they have
a $\Theta(1)$ space cost (Theorem~\ref{th-vertex-listing}).

Since we may suppose that the simple compact representation of $G$ is sorted,
line~3 can be achieved in $O(m)$ time.
The number of vertices in $G'$ is in $\Theta(\frac{m}{K})$ and it may be a
clique, thus the space needed for $G$ is in $\Theta((\frac{m}{K})^2)$.

Finally, the overall time complexity is in $O\left(m.K + m\cdot\frac{m}{K}\right)$.
The optimal is attained with $K$ in $\Theta(\sqrt{m})$, leading to the announced time
complexity (which is tight from Lemma~\ref{lem-bound}). The space complexity
then is $\Theta((\frac{m}{K})^2) = \Theta(m)$.
\end{proof}

Again, this result has a significant space cost: it needs the adjacency matrix of $G$,
and, even then, it needs $\Theta(m)$ additional
space. Moreover, it relies on the use of a parameter, $K$, which may be difficult
to choose in practice: though Theorem~\ref{th-ayz-listing} says that it must be in
$\Theta(\sqrt{m})$, this makes little sense when one considers a given graph.
We discuss further this issue in Section~\ref{sec-exp}.

\subsubsection*{The {\em forward} fast algorithm \cite{schank05finding,schank05findingWEA}.}

In \cite{schank05finding,schank05findingWEA}, the authors propose another
algorithm with optimal time complexity and a $\Theta(m)$ cost, while needing
only a simple compact representation of $G$. We now present it
in detail. We give a new
proof of the correctness and complexity of this algorithm, in order to be able
to extend it in the next sections (in particular in Section~\ref{sec-pl}).

\algorithm{forward}{Lists all the triangles in a graph \cite{schank05finding,schank05findingWEA}.}{
any simple compact representation of $G$}{
all the triangles in $G$}{
1. number the vertices with an injective function $\eta()$\\
\myindent\myindent such that $d(u)>d(v)$ implies $\eta(u)<\eta(v)$ for all $u$ and $v$\\
2. let $A$ be an array of $n$ sets initially equal to $\emptyset$\\
3. for each vertex $v$ taken in increasing order of $\eta()$:\\
\myindent 3a. for each $u \in N(v)$ with $\eta(u)>\eta(v)$:\\
\myindent\myindent 3aa. for each $w$ in $A[u] \cap A[v]$: output triangle $\{u,v,w\}$\\
\myindent\myindent 3ab. add $v$ to $A[u]$
}

\begin{theorem} \cite{schank05finding,schank05findingWEA}
Given any simple compact representation of $G$, it is possible to list
all its triangles in $\Theta(m^{\frac{3}{2}})$ time and
$\Theta(m)$ space;
\algoref{forward}\ achieves this.
\label{th-forward}
\end{theorem}
\begin{proof}
For all vertices $x$, let us denote
by $A(x)\ =\ \{y\in N(x),\ \eta(y)<\eta(x)\}$ the set of neighbors $y$ of $x$
with number $\eta(y)$ smaller than the one of $x$ itself. For any triangle
$t\ =\ \{a,b,c\}$ one can suppose without loss of generality that $\eta(c)<\eta(b)<\eta(a)$.
One may then discover $t$ by discovering that $c$ is in $A(a) \cap A(b)$.

This is what \algoref{forward}\ does. To show this it suffices to show that
$A[u] \cap A[v]\ =\ A(u)\cap A(v)$ when computed in line~3aa.

First notice that when one enters in the main loop (line~3), then
the set $A[v]$ contains all the vertices in $A(v)$.
Indeed, $u$ was previously treated by the main loop since $\eta(u)<\eta(v)$,
and during this lines~3 and~3ab ensure that it has been added to $A[v]$
(just replace $u$ by $v$ and $v$ by $u$ in the pseudocode). Moreover,
$A[v]$ contains no other element, and thus it is exactly $A(v)$ when
one enters the main loop.

Likewise, when entering the main loop for $v$, $A[u]$ is
not equal to $A(u)$, but it contains all the vertices $w$ such
that $\eta(w)<\eta(v)$ and that belong to
$A(u)$. Therefore, the intersections are equal:
$A[u] \cap A[v]\ =\ A(u)\cap A(v)$, and thus the algorithm is correct.

If we turn to the time complexity, first notice that line~1 can be
achieved in $\Theta(n\cdot\log(n))$ (and even in $\Theta(n)$) time
and $\Theta(n)$ space. This plays no role in the following.

Now, note that lines~3 and~3a are
nothing but a loop over all edges, thus in $\Theta(m)$. Inside the loop,
the expensive operation is the intersection computation. To obtain the
claimed complexity, it suffices to show that
both $A[u]$ and $A[v]$ contain $O(\sqrt{m})$ vertices (since each structure
$A[x]$ is trivially sorted by construction, this is sufficient to ensure
that the intersection computation is in $O(\sqrt{m})$).

For any vertex $x$, by definition of $A(x)$ and $\eta()$, $A(x)$ is included
in the set of neighbors of $x$ with degree at least $d(x)$. Suppose $x$
has $\omega(\sqrt{m})$ such neighbors: $|A(x)| \in \omega(\sqrt{m})$.
But all these vertices have degree
at least equal to the one of $x$, with $d(x)\ge|A(x)|$, and thus they
have all together
$\omega(m)$ edges, which is impossible. Therefore one must have
$|A(x)| \in O(\sqrt{m})$, and since $A[x] \subseteq A(x)$ this proves
the $O(m^{\frac{3}{2}})$ time complexity. This bound is tight from
Lemma~\ref{lem-bound}.

The space complexity is obtained when one notices that each edge induces
a $\Theta(1)$ space in $A$, leading to a global space in $\Theta(m)$.
\end{proof}

Compared to \algoref{ayz-listing}, this algorithm has several advantages
(although it has the same asymptotic time and space complexities).
It is very simple and easy to implement, which also implies, as shown
in \cite{schank05finding,schank05findingWEA}, that it is very efficient
in practice. Moreover, it does not have the drawback of depending on
a parameter $K$, central in \algoref{ayz-listing}.
Finally, we show in the next sections that it may be slightly modified
to obtain a $\Theta(n)$ space complexity (Section~\ref{sec-listing-compact}),
and that even better performances can be
proved if one considers power-law graphs (Section~\ref{sec-exp}).

\subsection{Time-optimal {\em compact} algorithms for sparse graphs.}
\label{sec-listing-compact}

This section is devoted to listing algorithms that have very low space requirements,
both in terms of the given representation of $G$ and in terms of the additional
space needed. We will obtain two algorithms reaching a $\Theta(n)$ space cost
while needing only a simple compact representation of $G$, and in optimal
$\Theta(m^{\frac{3}{2}})$ time.

\subsubsection*{A compact version of \algoref{forward}.}
\label{sec-compact-forward}

Thanks to the proof we gave of Theorem~\ref{th-forward}, it is now
easy to modify \algoref{forward} in order to improve significantly
its space complexity. This leads to the following result.

\algorithm{compact-forward}{Lists all the triangles in a graph.}{
any simple compact representation of $G$}{
all the triangles in $G$}{
1. number the vertices with an injective function $\eta()$\\
\myindent\myindent such that $d(u)>d(v)$ implies $\eta(v)>\eta(u)$ for all $u$ and $v$\\
2. sort the simple compact representation according to $\eta()$\\
3. for each vertex $v$ taken in increasing order of $\eta()$:\\
\myindent 3a. for each $u \in N(v)$ with $\eta(u)>\eta(v)$:\\
\myindent\myindent 3aa. let $u'$ be the first neighbor of $u$, and $v'$ the one of $v$\\
\myindent\myindent 3ab. while there remains untreated neighbors of $u$ and $v$ and $\eta(u')<\eta(v)$ and $\eta(v')<\eta(v)$:\\
\myindent\myindent\myindent 3aba. if $\eta(u')<\eta(v')$ then set $u'$ to the next neighbor of $u$\\
\myindent\myindent\myindent 3abb. else if $\eta(u')>\eta(v')$ then set $v'$ to the next neighbor of $v$\\
\myindent\myindent\myindent 3abc. else:\\
\myindent\myindent\myindent\myindent 3abca. output triangle $\{u,v,u'\}$\\
\myindent\myindent\myindent\myindent 3abcb. set $u'$ to the next neighbor of $u$\\
\myindent\myindent\myindent\myindent 3abcc. set $v'$ to the next neighbor of $v$\\
}

\begin{theorem}
Given any simple compact representation of $G$, it is possible to list all its
triangles in $\Theta(m^{\frac{3}{2}})$ time and $\Theta(n)$ space;
\algoref{compact-forward}\ achieves this.
\label{th-compact-forward}
\end{theorem}
\begin{proof}
Recall that, as explained in the proof of Theorem~\ref{th-forward}, when one
computes the intersection of $A[v]$ and $A[u]$ (line~3aa of \algoref{forward}),
$A[v]$ is the set of neighbors of $v$ with number lower than $\eta(v)$, and
$A[u]$ is the set of neighbors of $u$ with number lower than $\eta(v)$. If the
adjacency structures encoding the neighborhoods are sorted according to
$\eta()$, we then have that $A[v]$ is nothing but the beginning of $N(v)$,
truncated when we reach a vertex $v'$ with $\eta(v')>\eta(v)$. Likewise, $A[u]$
is $N(u)$ truncated at $u'$ such that $\eta(u')>\eta(v)$.

\algoref{compact-forward}\ uses this. Indeed, lines~3ab to~3abcc are nothing
but the computation of the intersection of $A[v]$ and $A[u]$, which are supposed
to be stored at the beginning of the adjacency structures, which is done in
line~2. All this has no impact on the asymptotic time cost, and now the $A$
structure does not have to be explicitly stored.

Notice now that line~1 has a $O(n\cdot\log(n))$ time and $\Theta(n)$ space 
cost. Moreover, sorting the simple compact representation of $G$ (line~2) is in
$O(m\cdot \log(n))$ time and $\Theta(1)$ space. These time complexities play
no role in the overall complexity, but the space complexities induce a
$\Theta(n)$ space cost for the overall algorithm.

Finally, the time cost is the same as the one of \algoref{forward}, and
the space cost is in $\Theta(n)$.
\end{proof}

In practice, this result means that one may encode vertices by integers,
with the property that this numbering goes from highest degree vertices
to lowest ones, then store the graph in a simple compact representation,
sort it, and compute the triangles using \algoref{compact-forward}. In
such a framework, it is important to notice that the algorithm runs
in $\Theta(1)$ space, since line~1, responsible for the $\Theta(n)$
cost, is unnecessary. On the other hand, if one wants to keep the
original numbering of the vertices, then one has to store the function $\eta()$
and renumber the vertices back after the triangle computation. This has
a $\Theta(n)$ space cost (and no significant time cost). Going further,
if one wants to restore the initial order inside the simple sorted representation,
then one has to sort it back if it was sorted before the computation, and
even to store a copy of it (then in $\Theta(m)$ space) if it was unsorted.

\subsubsection*{A new algorithm.}

The algorithms discussed until now basically rely on the fact that they
avoid considering each pair of neighbors of high degree vertices,
which would have a prohibitive cost. They do so by managing low degree
vertices first, which has the consequence that most edges involved in
the highest degrees have already been treated when the algorithm comes
to these vertices.
Here we take a quite different approach. First we
design an algorithm able to efficiently list the triangles of high degree
vertices. Then, we use it in an algorithm similar to \algoref{ayz-listing},
but that both avoids adjacency matrix representation,
and reaches a $\Theta(n)$ space cost.

First note that we already have an algorithm listing all the triangles
containing a given vertex $v$, namely \algoref{vertex-listing}\ \cite{itai78finding}.
This algorithm is in $\Theta(1)$ space, but it is unefficient
on high degree vertices, since it needs $\Theta(d(v)^2)$ time.
Our improved listing algorithm relies on an equivalent to \algoref{vertex-listing}\ 
that avoids this.

\algorithm{new-vertex-listing}{Lists all the triangles containing a given vertex.}{
any simple compact representation of $G$, and a vertex $v$}{
all the triangles to which $v$ belongs}{
1. create an array $A$ of $n$ booleans and set them to \false\\
2. for each vertex $u$ in $N(v)$, set $A[u]$ to \true\\
3. for each vertex $u$ in $N(v)$:\\
\myindent 3a. for each vertex $w$ in $N(u)$:\\
\myindent\myindent 3aa. if $A[w]$ then output $\{v,u,w\}$
}

\begin{lemma}
Given any simple compact representation of $G$, it is posible to list all
its triangles containing a given vertex $v$ in
$\Theta(m)$ (optimal) time and $\Theta(n)$ space; \algoref{new-vertex-listing}\ achieves
this.
\label{lem-new-vertex-listing}
\end{lemma} 
\begin{proof} 
One may see \algoref{new-vertex-listing}\ as a way to use the adjacency
matrix of $G$ without explicitely storing it: the array $A$ is nothing
but the $v$-th line of the adjacency-matrix. It is constructed in $\Theta(n)$
time and space (lines~1 and~2). Then one can test for any edge $(v,u)$ in $\Theta(1)$
time and space. The loop starting at line $3$ takes any edge containing one
neighbor $u$ of $v$ and tests if its other end ($w$ in the pseudo-code) is linked to
$v$ using $A$, in $\Theta(1)$ time and space. This is sufficient to find all
the triangles containing $v$. Since this number of edges is
bounded by $2\cdot m$ (one may actually obtain an equivalent algorithm by replacing
lines~3a and~3aa by a loop over all the edges), we obtain that the algorithm
is in $O(m)$ time and $\Theta(n)$ space.

The obtained time complexity is optimal since $v$
may belong to $\Theta(m)$ triangles.
\end{proof}

\algorithm{new-listing}{Lists all the triangles in a graph.}{
any sorted simple compact representation of $G$, and an integer $K$}{
all the triangles in $G$}{
1. for each vertex $v$ in $V$:\\
\myindent1a. if $d(v) > K$ then, using \algoref{new-vertex-listing}:\\
\myindent\myindent1aa. output all triangles $\{v,u,w\}$ such that $d(u)>K$, $d(w)>K$ and $v>u>w$\\
\myindent\myindent1ab. output all triangles $\{v,u,w\}$ such that $d(u)>K$, $d(w)\le K$ and $v>u$\\
\myindent\myindent1ac. output all triangles $\{v,u,w\}$ such that $d(u)\le K$, $d(w)>K$ and $v>w$\\
2. for each edge $(v,u)$ in $E$:\\
\myindent2a. if $d(v) \le K$ and $d(u) \le K$ then:\\
\myindent\myindent2aa. if $u<v$ then output all triangles containing $(u,v)$ using \algoref{edge-listing}
}

\begin{theorem}
Given any sorted simple compact representation of $G$, it is possible to list all its
triangles in $\Theta(m^{\frac{3}{2}})$ time and $\Theta(n)$ space; \algoref{new-listing}\ achieves this
if one takes $K \in \Theta(\sqrt{m})$.
\label{th-new-listing}
\end{theorem}
\begin{proof}
Similarily to the proof we gave of Theorem~\ref{th-ayz-listing}, let us first study
the complexity of \algoref{new-listing} as a function of $K$. For each vertex $v$
with $d(v)> K$, one lists the number of triangles containing $v$ in
$\Theta(m)$ time and $\Theta(n)$ space (Lemma~\ref{lem-new-vertex-listing}) (the
conditions in lines~1aa to 1ac, as well as the one in line~2aa, only serve to ensure
that each triangle is listed exactly once).
Then, one lists the triangles containing edges whose extremities
are of degree at most $K$; this is done by \algoref{edge-listing}\ in
$\Theta(K)$ time and $\Theta(1)$ space for each edge, thus a total in
$O(m\cdot K)$ time and $\Theta(1)$ space.

Finally, the space complexity of the whole algorithm is independent
of $K$ and is in $\Theta(n)$, and its time complexity is in
$O(\frac{m}{K}\cdot m + m\cdot K)$ time, since there are
$O(\frac{m}{K})$ vertices with degree larger than $K$.
In order to minimize this, we now take $K$ in $\Theta(\sqrt{m})$,
which leads to the announced time complexity.
\end{proof}

Theorems~\ref{th-compact-forward} and~\ref{th-new-listing} improve
Theorems~\ref{th-ayz-listing} and~\ref{th-forward} since they show that the
same (optimal) time-complexity may be achieved in space $\Theta(n)$ rather than $\Theta(m)$.
Moreover, this is space-optimal for pseudo-listing if one wants to keep the result
in memory (the result itself is in $\Theta(n)$), which is generally the case (for
clustering coefficient computations, for instance).

Note however that it is still unknown wether there exist algorithms with
time complexity in $\Theta(m^{\frac{3}{2}})$ but with $o(n)$ space requirements. We
saw that \edgeiterator\ achieves $\Theta(m\cdot d_{\max}) \subseteq O(m\cdot n)$
time and $\Theta(1)$ space complexities (Theorem~\ref{th-edge-listing}),
while needing only a sorted simple compact representation of $G$. If we
suppose that the representation uses adjacency {\em arrays}, we obtain now
the following stronger (if $d_{\max} \in \Omega(\sqrt{m\cdot\log(n)})$) result.

\begin{corollary}
Given the adjacency array representation of $G$, it is possible to list all its
triangles in $O(m^{\frac{3}{2}}\sqrt{\log(n)})$ time and $\Theta(1)$ space;
\algoref{new-listing}\ achieves this if one takes $K \in \Theta(\sqrt{m\cdot \log(n)})$.
\label{cor-listing-1}
\end{corollary}
\begin{proof}
Let us first sort the arrays in $O(m\cdot \log(n))$ time and $\Theta(1)$ space.
Then, we change \algoref{new-vertex-listing}\ by removing the
use of $A$ and replace line~3aa by a dichotomic search for $w$ in $N(u)$,
which has a cost in $O(\log(n))$ time and $\Theta(1)$ space.
Now if \algoref{new-listing}\ uses this modified version of \algoref{new-vertex-listing}, then
it is in $\Theta(1)$ space and $O(\frac{m}{K}\cdot m\cdot \log(n) + m\cdot K)$
time. The optimal value for $K$ is then in $\Theta(\sqrt{m\cdot \log(n)})$, leading to the announced
complexity.
\end{proof}

\section{The case of power-law graphs.}
\label{sec-pl}

Until now, several results (including ours) took advantage of the fact that most
large graphs met in practice are sparse; designing algorithms with complexities
expressed in term of $m$ rather than $n$ then leads to significant improvements.

Going further, it has been observed since several years that most large graphs
met in pratice also have another important characteristic in common: their degrees
are very heterogeneous. More precisely, in most cases, the vast majority of
vertices have a very low degree while some have a huge degree.
This is often captured by the fact that the degree distribution, \ie\ the proportion
$p_k$ for each $k$ of vertices of degree $k$, is well fitted by a power-law:
$p_k \sim k^{-\alpha}$ for an exponent $\alpha$ generally between $2$ and $3$. See
\cite{watts98collective,brandes05lncs,albert02statistical,milo02network,yeger04network,eubank04structural}
for extensive lists of cases in which this property was observed\,\footnote{Note that
if $\alpha$ is a constant then $m$ is in $\Theta(n)$. It may however depend on $n$,
and should be denoted by $\alpha(n)$. In order to keep the notations simple, we
do not use this notation, but one must keep this in mind.}.

We will see that several algorithms proposed in previous section have provable
better performances on such graphs than on general (sparse) graphs.

\medskip

Let us first note that there are several ways to model real-world power-law
distributions; see for
insance~\cite{Dorogovtsev2001Comment,Cohen2001Reply}. We use here one
of the most simple and classical ones, namely {\em continuous power-laws};
choosing one of the others would lead to similar results.
In such a distribution, $p_k$ is taken to be equal
to $\int_k^{k+1} C x^{-\alpha} \mathrm{d}x$, where $C$ is the
normalization constant\,\footnote{One may also choose $p_k$ proportional
to $\int_{k-\frac{1}{2}}^{k+\frac{1}{2}} x^{-\alpha}\mathrm{d}x$. Choosing any of this kind
of solutions has little impact on the obtained results, see~\cite{Cohen2001Reply} and the
proofs we present in this section.
}. This ensures that $p_k$ is
proportional to $k^{-\alpha}$ in the limit where $k$ is large.
We must moreover ensure that the sum of the $p_k$ is equal to $1$:
$\sum_{k=1}^\infty p_k = \int_1^{\infty} C\ x^{-\alpha} \mathrm{d}x 
= C\ \frac{1}{\alpha-1} = 1$. We obtain $C = \alpha-1$, and finally
$p_k = \frac{1}{\alpha - 1} \cdot \int_k^{k+1} x^{-\alpha} \mathrm{d}x = k^{-\alpha+1} - (k+1)^{-\alpha+1}$.

Finally, when we will talk about power-law graphs in the following, we will
refer to graphs in which the proportion of vertices of degree $k$ is
$p_k = k^{-\alpha+1} - (k+1)^{-\alpha+1}$.

\begin{theorem}
Given any simple compact representation of a power-law graph $G$ with exponent $\alpha$, it
is possible to list all its triangles in
$O(m\cdot n^{\frac{1}{\alpha}})$ time
and $\Theta(n)$ space; \algoref{new-listing}\ achieves this if one takes
$K \in \Theta(n^{\frac{1}{\alpha}})$, and \algoref{compact-forward}\ achieves this
too.
\label{th-pl}
\end{theorem}
\begin{proof}
Let us denote by $n_K$ the number of vertices of degree larger than
or equal to $K$.
In a power-law graph with exponent $\alpha$, this
number is given by: $\frac{n_K}{n} = \sum_{k=K}^{\infty} p_k$. We have
$\sum_{k=K}^{\infty} p_k
= 1-\sum_{k=1}^{K-1} p_k
= 1 - (1-K^{-\alpha+1})
= K^{-\alpha+1}$.
Therefore $n_K = n\cdot K^{-\alpha+1}$.

Let us first prove the result concerning \algoref{new-listing}.
As already noticed in the proof of Theorem~\ref{th-new-listing}, its space
complexity does not depend on $K$, and it is $\Theta(n)$. Moreover, its
time complexity is in $O(n_K\cdot m + m\cdot K)$.
The value of $K$ that minimizes this
is in $\Theta(n^{\frac{1}{\alpha}})$,
and the result for \algoref{new-listing}\ follows.

Let us now consider the case of \algoref{compact-forward}. The space
complexity was already proved for Theorem~\ref{th-compact-forward}.
The time complexity is the same as the one for \algoref{forward}, and
we use here the same notations as in the proof of Theorem~\ref{th-forward}.
Recall that the vertices are numbered by
decreasing order of their degrees.

Let us study the complexity of the intersection computation (line~3aa in
\algoref{forward}). It is in $\Theta(|A[u]| + |A[v]|)$. Recall that,
at this point of the algorithm, $A[v]$ is nothing but the set of neighbors
of $v$ with number lower than the one of $v$ (and thus of degree at
least equal to $d(v)$). Therefore, $|A[v]|$ is bounded both by $d(v)$
and the number of vertices of degree at least $d(v)$, \ie\ $n_{d(v)}$.
Likewise, $|A[u]|$ is bounded by $d(u)$ and by $n_{d(v)}$, since $A[u]$
is the set of neighbors of $u$ with degree at least equal to $d(v)$.
Moreover, we have $\eta(u)>\eta(v)$ (line~3a of \algoref{forward}), and so
$|A[u]| \le d(u) \le d(v)$. Finally, both $|A[u]|$ and $|A[v]|$ are
bounded by both $d(v)$ and $n_{d(v)}$, and the intersection computation
is in $O(d(v)+n_{d(v)})$.

Like above, let us compute the value $K$ of $d(v)$ such that these
two bounds are equal. We obtain $K = n^{\frac{1}{\alpha}}$. Then, the computation
of the intersection is in $O(K+n_K) = O(n^{\frac{1}{\alpha}})$, and since the
number of such computations is bounded by the number of edges (lines~3
and~3a of \algoref{forward}), we obtain the announced complexity.
\end{proof}

This result improves significantly the known bounds, as soon as $\alpha$
is large enough. This holds in particular for typical cases met in practice,
where $\alpha$ often is between $2$ and $3$ \cite{brandes05lncs,albert02statistical}. It may be seen as
an explanation of the fact that \algoref{forward}\ has very good performances
on graphs with heterogeneous degree distributions, as shown experimentally
in \cite{schank05finding,schank05findingWEA}.

\medskip

One may also use this approach to improve
\algoref{ayz-pseudo-listing}\ and \algoref{ayz-listing} in the case of
power-law graphs as follows.

\begin{corollary}
Given any simple compact representation of a power-law graph $G$ with exponent $\alpha$
and its adjacency matrix, it
is possible to solve pseudo-listing, counting and finding on $G$ in
$O(n^{\frac{\omega\cdot\alpha+\omega}{\omega\cdot\alpha-\omega+2}})$ time
and $\Theta(n^{\frac{2\cdot\alpha +2}{\omega\cdot\alpha - \omega + 2}})$ space;
\algoref{ayz-pseudo-listing}\ achieves this if one takes
$K$ in $\Theta(n^{\frac{\omega-1}{\omega\cdot\alpha - \omega + 2}})$.
\label{cor-ayz-pseudo-listing}
\end{corollary}
\begin{proof}
With the same reasoning as the one in the proof of Theorem~\ref{th-ayz-pseudo-listing},
one obtains that the algorithm runs in $O(n\cdot K^2 + (n_K)^{\omega})$ where $n_K$
denotes the number of vertices of degree larger than $K$. As explained in the proof of
Theorem~\ref{th-pl}, this is $n_K = n\cdot K^{-\alpha+1}$. Therefore, the best $K$ is
such that $n\cdot K^2$ is in $\Theta( n^{\omega}\cdot K^{\omega\cdot(1-\alpha)})$.
Finally, $K$ must be in $n^{\frac{1-\omega}{\omega\cdot(1-\alpha) - 2}}$. One then
obtains the announced time complexity. The space complexity is bounded by the space
needed to construct the adjacency matrix between the vertices of degree at most $K$,
thus it is $(n_K)^2$, and the result follows.
\end{proof}

If the degree distribution of $G$ follows a power law with exponent $\alpha = 2.5$
(typical for internet graphs \cite{brandes05lncs,albert02statistical}) then this result says that \algoref{ayz-pseudo-listing}\ 
reaches a $O(n^{1.5})$ time and $O(n^{1.26})$ space complexity.
If the exponent is larger, then the complexity
is even better. Note that one may also obtain tighter bounds in terms of $m$ and $n$,
for instance using the fact that \algoref{ayz-pseudo-listing}\ has running time in
$\Theta(m\cdot K + (n_K)^{\omega})$ rather than $\Theta(n\cdot K^2 + (n_K)^{\omega})$
(see the proofs of Theorem~\ref{th-ayz-pseudo-listing} and Corollary~\ref{cor-ayz-pseudo-listing}).
We do not detail this here because the obtained results are quite technical and follow
immediately from the ones we detailed.

\begin{corollary}
Given any simple compact representation of a power-law graph $G$ with exponent $\alpha$
and its adjacency matrix, it is possible to
list all its triangles in $\Theta(m\cdot n^{\frac{1}{\alpha}})$ time
and $\Theta(n^{\frac{2}{\alpha}})$ space; \algoref{ayz-listing}\ achieves this if one takes
$K$ in $\Theta(n^{\frac{1}{\alpha}})$.
\label{cor-ayz-listing}
\end{corollary}
\begin{proof}
The time complexity of \algoref{ayz-listing}\ is in $\Theta(m\cdot K + m\cdot n_K)$.
The $K$ minimizing this is such that $K \in \Theta(n_K)$, which is the same
condition as the one in the proof of Theorem~\ref{th-pl}; therefore we reach the
same time complexity. The space complexity is bounded by the size of the adjacency
matrix, \ie\ $\Theta((n_K)^2)$. This leads to the announced complexity.
\end{proof}

Notice that this result implies that, for some reasonable values of $\alpha$
(namely $\alpha>2$) the space complexity is in $o(n)$. This however is of theoretical
interest only: it relies on the use of both the adjacency matrix and a simple
compact representation of $G$, which is unfeasable in practice for large graphs.

\medskip

Finally, the results presented in this section show that one may use
properties of most large graphs met in practice (here, their heterogeneous
degree distribution), to improve results known on the general case (or on the
sparse graph case).
As we discuss further in Section~\ref{sec-conclu},
using such properties in the design of algorithms
is a promising direction for algorithmic research
on very large graphs met in practice.

We note however that we have no
lower bound for the complexity of triangle listing with the assumption that
the graph is a power-law one (which we had for general and sparse
graphs); actually, we do not even have a proof of the fact that the
given bound is tight for the presented algorithms. One may therefore
prove that they have even better performance (or that the bound is
tight), and algorithms faster
than the ones presented here may exist (for power-law graphs).

\section{Experimental evaluation.}
\label{sec-exp}

In \cite{schank05finding,schank05findingWEA}, the authors present a wide set of
experiments on both real-world
complex networks and some generated using various models, to evaluate experimentally
the known algorithms. They focus on \vertexiterator, \edgeiterator, \algoref{forward},
and \algoref{ayz-listing}, together with their counting and pseudo-listing variants
(they compute clustering coefficients).
They also study variants of these algorithms using for instance hashtables and balanced
trees. These variants have the same worst case asymptotic complexities but one may guess
that they would run faster than the original algorithms, for several reasons we do not
detail here. Matrix approaches are considered as too intricate
to be used in practice.

The overall conclusion of their extensive experiments is that \algoref{forward}\ performs
best on real-world (sparse and power-law) graphs: its asymptotic time is optimal and
the constants involved in its implementation
are very small. Variants, which need more subtle data structure, actually fail in
performing better in most cases (because of the overhead induced by the management
of these structures).

\medskip

In order to integrate our contribution in this context and have a precise idea of
the behavior of the discussed algorithms in practice, we also performed a wide set of
experiments\,\footnote{Optimized implementations are provided at \cite{progurl}.}.
They confirm that \algoref{forward}\ is very fast and outperforms
classical approaches significantly. They also show that, even in the cases where
available memory is sufficient for this algorithm, it is outperformed by
\algoref{compact-forward}\ because it avoids management of additional
data structures.

Note that \algoref{new-listing}, just like \algoref{ayz-pseudo-listing}\ and
\algoref{ayz-listing}, suffers from a serious drawback: it relies on the choice
of a relevant value for $K$, the maximal degree above which vertices are
considered as having a high degree. Though in theory this is not a problem, in
practice it may be quite difficult to determine the best value for $K$, \ie\ the
one that minimizes the execution time. It depends both on the machine running the
program and on the graph under concern.
One may evaluate the best $K$ in a preprocessing step
at running time, by measuring the time needed to perform the key steps of the algorithm
for various $K$. This can be done without changing the asymptotic complexity.
However, there is a much simpler way to choose $K$, with neglectible loss in
performance, which we discuss below. Until then, we suppose that we
were able to determine the best value for $K$.

With this best value given, the performances of \algoref{new-listing}\ are
similar to the ones of \algoref{forward}; its space requirements are much
lower, as predicted by Theorem~\ref{th-new-listing}. Likewise, \algoref{new-listing}\ 
speed is close to the one of \algoref{compact-forward}\ and it has the same
space requirements.

\medskip

It is important to notice that the use of compact algorithms, namely
\algoref{compact-forward}\ and \algoref{new-listing}, makes it possible
to manage graphs that were previously out of reach because of space
requirements. To illustrate this, we present now an experiment on a huge
graph which previous algorithms were unable to manage in our $8$~GigaBytes memory
machine. This experiment also has the advantage of being representative of
what we observed on a wide variety of instances.

\medskip

\begin{figure}[!h]
\hfill \includegraphics[scale=0.4]{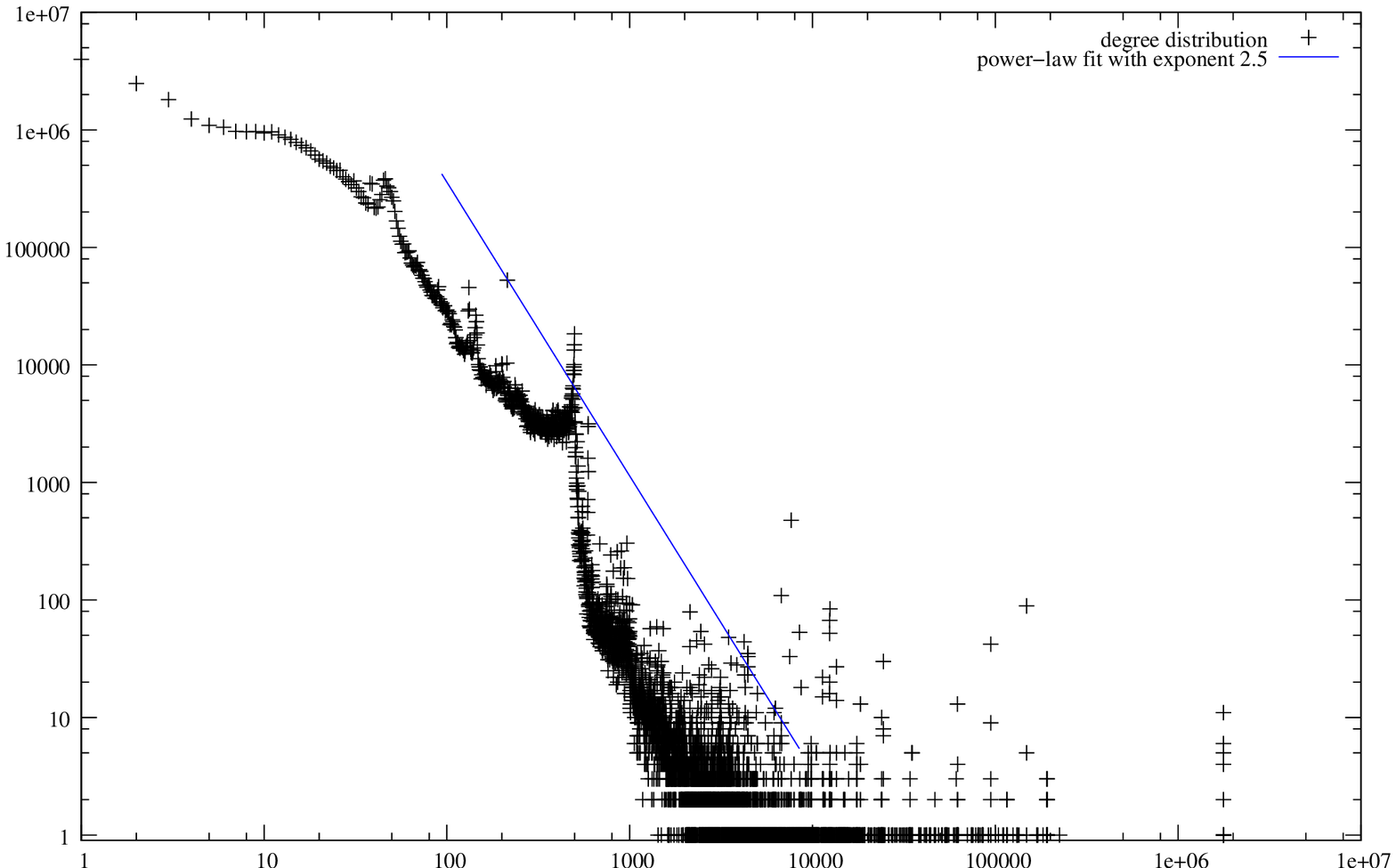}\ \ \ 
\hfill
\ \ \ \includegraphics[scale=0.4]{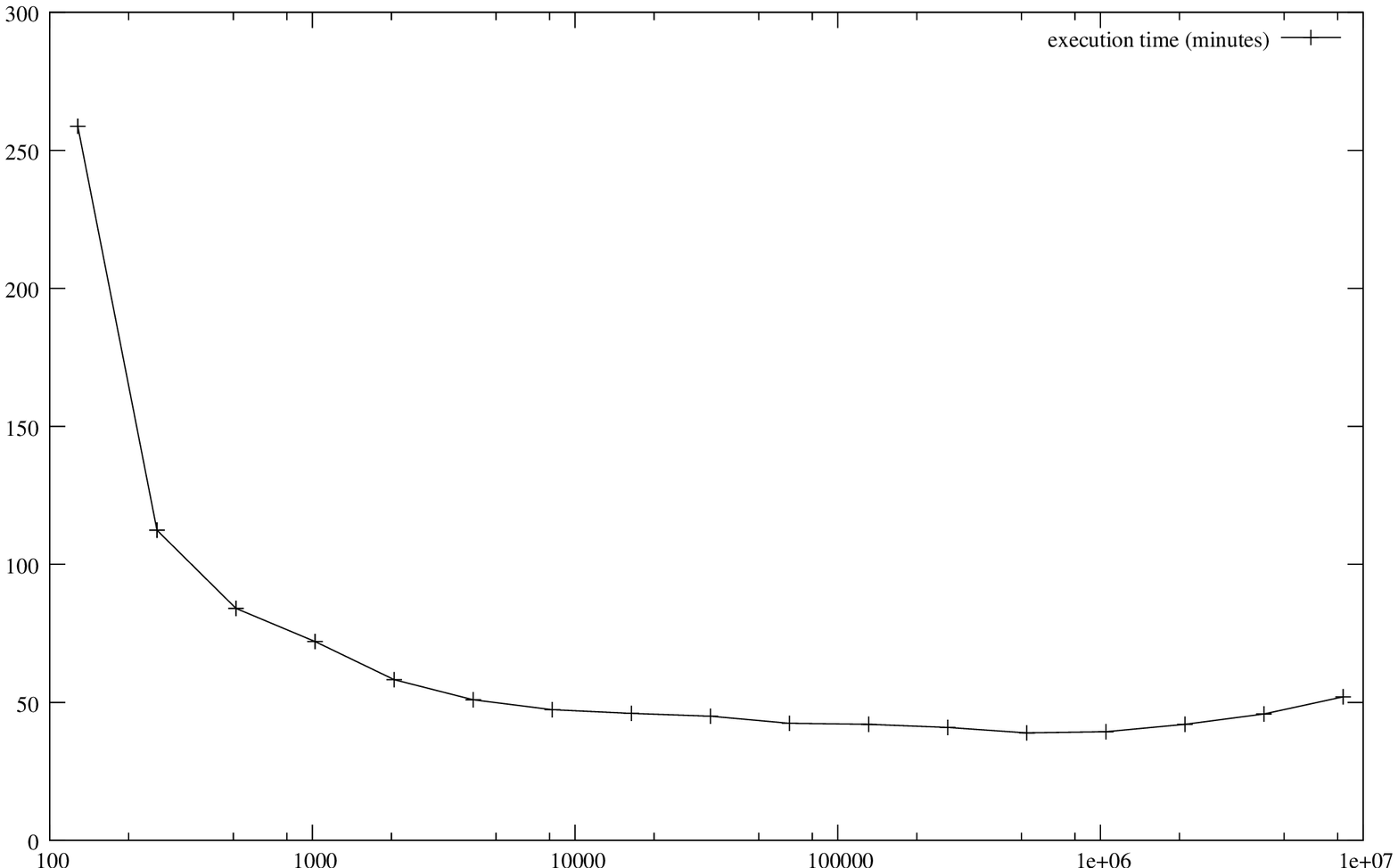} \hfill\\
\caption{Left: the degree distribution of our graph. Right: the execution time (in
minutes) as a function of the number of vertices considered as high degree ones.}
\label{fig-time}
\label{fig-deg-distrib}
\end{figure}

The graph we consider here is a {\em web} graph provided by the
{\em WebGraph} project \cite{webgraphurl}. It contains all the web pages in the {\tt .uk}
domain discovered during a crawl conducted from the 11-th of
july, 2005, at 00:51, to the 30-th at 10:56 using
{\em UbiCrawler} \cite{boldi04ubicrawler}.
It has $n=39,459,925$ vertices and $m=783,027,125$ (undirected) edges, leading to more
than $6$ GigaBytes of memory usage if stored
in (sorted) (uncompressed) adjacency arrays, each vertex being encoded in $4$ bytes as
an integer between $0$ and $n-1$. Its degree distribution is plotted in
Figure~\ref{fig-deg-distrib}, showing that the degrees are very heterogeneous
and reasonably well fitted by a power-law of exponent $\alpha=2.5$.
It contains $304,529,576$ triangles.

Let us insist on the fact that \algoref{forward}, as well as the ones based on
adjacency matrices, are unable to manage this graph on our $8$~GigaBytes
memory machine.
Instead, and despite the fact that it is quite slow, \edgeiterator, with
its $\Theta(1)$ space complexity, can handle this. It took approximately $41$ hours to solve
pseudo-listing on this graph with this algorithm on our machine.

\algoref{compact-forward}\ achieves much better results: it took approximately
$20$ minutes. Likewise, \algoref{new-listing}\ took around $45$ minutes (depending
on the value of $K$).
This is probably close to what \algoref{forward}\ would achieve in $16$~GigaBytes
of central memory.

\medskip

We plot in
Figure~\ref{fig-time}~(right) the running time of \algoref{new-vertex-listing}
as a function of the number of vertices with degree larger than $K$,
for varying values of $K$. Surprisingly enough,
this plot shows clearly that the time performance increases
drastically as soon as a few vertices are considered as high degree ones.
This may be seen as a consequence of the fact that \edgeiterator\ is very
efficient when the maximal degree is bounded; managing high degree vertices
efficiently with \algoref{new-vertex-listing}\ and then the low degree ones
with \edgeiterator\ therefore leads to good performances. In other words, the
few high degree vertices (which may be observed on the degree distribution
plotted in Figure~\ref{fig-deg-distrib}) are responsible for the low
performance of \edgeiterator.

When $K$ decreases, the number of vertices with degree larger than $K$ increases,
and the performances continue to be better and better for a while. They reach
a minimal running time,
and then the running time grows again. The other important point here is that
this growth is very slow, and thus the
performance of the algorithm remains close to its best for a wide range of values
of $K$. This implies that, with any reasonable guess for $K$, the algorithm performs
well.

\section{Conclusion.}
\label{sec-conclu}

In this contribution, we gave a detailed survey of existing results on
triangle problems, and we completed them in two directions. First, we
gave the space complexity of each previously known algorithm. Second,
we proposed new algorithms that achieve both optimal time complexity and
low space needs. Taking space requirements into account is a key issue in
this context, since this currently is the bottleneck for triangle problems
when the considered graphs are very large. This is discussed on a practical
case in Section~\ref{sec-exp}, where we show that our compact algorithms
make it possible to handle cases that were previously out of reach.

Another significant contribution of this paper is the analysis of
algorithm performances on power-law graphs (Section~\ref{sec-pl}),
which model a wide variety
of very large graphs met in practice. We were able to show that, on
such graphs, several algorithms have better performance than in the
general (sparse) case.

\noindent
Finally, the current state of the art concerning triangle problems,
including our new results, may be summarized as follows:
\vspace{-2mm}
\begin{itemize}\setlength{\itemsep}{-1mm}
\item
except the fact that pseudo-listing may have a $\Theta(n)$ space overhead
(depending on the underlying algorithm), there is no known difference in
time and space complexities between finding, counting, and pseudo-listing;
\item
the fastest known algorithms for these three problems rely on
matrix product and are in $O(n^{2.376})$ time and $\Theta(n^2)$ space
(Theorem~\ref{th-matrix}), or in $O(m^{1.41})$ time and
$O(m^{1.185})$ space (Theorem~\ref{th-ayz-pseudo-listing});
however, no lower bound better than the trivial $\Omega(m)$ one
is known for the time complexity of these problems;
\item
the other known algorithms rely on solutions to the listing problem
and have the same performances as on this problem; they are
slower than matrix approaches but need less space;
\item
listing can be solved in $\Theta(n^3)$ or $\Theta(n\cdot m)$ (optimal in the general case) time and
$\Theta(1)$ (optimal) space (Theorems~\ref{th-direct},~\ref{th-vertex-listing}
and~\ref{th-edge-listing}); this can be achieved from a sorted simple compact
representation of the graph;
\item
listing may also be solved in $\Theta(m^{\frac{3}{2}})$ (optimal in the general and
sparse cases) time and $\Theta(n)$ space (Theorems~\ref{th-new-listing}
and~\ref{th-compact-forward}), still from a simple compact representation of the graph;
this is much better for sparse graphs;
\item
in the case of power-law graphs, it is possible to prove better complexities,
leading to $O(m\cdot n^{\frac{1}{\alpha}})$ time and $\Theta(n)$ space solutions,
where $\alpha$ is the exponent of the power-law (Theorem~\ref{th-pl});
\item 
in practice, it is possible to obtain very good performances (both concerning
time and space needs) using \algoref{new-listing}\ and \algoref{compact-forward}.
\end{itemize}
We detailed several other results, but they are weaker (they need the adjacency
matrix of the graph in input and/or have higher complexities) than these ones.


\medskip

\noindent
This contribution also opens a set of questions for further research, most
of them related to the tradeoff between space and time efficiency. Let
us cite for instance:
\vspace{-2mm}
\begin{itemize}\setlength{\itemsep}{-1mm}
\item can matrix approaches be modified in order to induce less space complexity?
\item is listing feasable in $o(n)$ space, while still
in optimal time $\Theta(m^{\frac{3}{2}})$?
\item is it possible to design a listing algorithm with complexity $o(m\cdot n^{\frac{1}{\alpha}})$
time and $o(n)$ space for power-law graphs with exponent $\alpha$? what
is the optimal time complexity in this case?
\end{itemize}

It is also important to notice that other approaches exist, based for instance on streaming
algorithmics (avoiding to store the graph in central memory)
\cite{henzinger98computing,bar02reduction,jowhari05new} and/or approximate
algorithms \cite{schank04approximating,jowhari05new,shapira05homomorphisms}, and/or
various methods to compress the graph \cite{boldi04www,boldi04dcc}.
These approaches are very promising for graphs even larger than the
ones considered here, in particular the ones that do not fit in
central memory.

Another interesting approach would be to express the complexity of
triangle algorithms in terms of the number of triangles in the graph
(and of its size). Indeed, it may be possible to achieve much better
performance for listing algorithms if the graph contains few triangles.
Likewise, it is reasonable to expect that triangle listing, but also
pseudo-listing and counting, may perform poorly if there are many
triangles in the graph. The finding problem, on the contrary, may
be easier on graphs having many triangles. To our knowledge, this
direction has not yet been explored.

Finally, the results we present in Section~\ref{sec-pl} take
advantage of the fact that most very large graphs considered in
practice may be approximed by power-law graphs. It is not the first time
that algorithms for triangle problems use underlying
graph properties to get improved performance.
For instance, results on planar graphs are provided
in \cite{itai78finding}, and results using arboricity in \cite{chiba85arboricity,alon97finding}.
It however appeared quite recently that many large graphs met in practice have
some nontrivial (statistical) properties in common, and using these properties
in the design of efficient algorithms still is at its very beginning.
We consider this as a key direction for further research.

\small

\bigskip
\noindent
{\bf Acknowledgments.}
I warmly thank Fr\'ed\'eric Aidouni, Michel Habib, Vincent Limouzy,
Cl\'emence Magnien, Thomas Schank and Pascal Pons for helpful comments and references.
I also thank Paolo Boldi from the WebGraph project \cite{webgraphurl}, who
provided the data used in Section~\ref{sec-exp}.
This work was partly funded by the MetroSec (Metrology of the Internet for Security)
\cite{metrosecurl} and
PERSI (Programme d'\'Etude des R\'eseaux Sociaux de l'Internet) \cite{persiurl}
projects.

\bibliographystyle{plain}
\bibliography{biblio}

\end{document}